\documentstyle[twocolumn,pra,aps]{revtex}
\def\begmc{}
\def\endmc{}

\begin{document}
\title{Quantum limits in interferometric measurements}
\author{M.T. Jaekel $^a$ and S. Reynaud $^b$}
\address{($a$) Laboratoire de Physique Th\'{e}orique de l'Ecole Normale
Sup\'{e}rieure \thanks{%
Unit\'e propre du Centre National de la Recherche Scientifique associ\'{e}e
\`{a} l'Ecole Normale Sup\'{e}rieure et \`{a} l'Universit\'{e} de Paris-Sud.}
24 rue Lhomond F-75231 Paris Cedex 05\\
($b$) Laboratoire de Spectroscopie Hertzienne \thanks{%
Unit\'e de l'Ecole Normale Sup\'{e}rieure et de l'Universit\'{e} Pierre et
Marie Curie, associ\'{e}e au Centre National de la Recherche Scientifique.}
Universit\'{e} Pierre et Marie Curie 4 place Jussieu F-75252 Paris Cedex 05}
\date{{\sc Europhysics Letters} {\bf 13} 301-306 (1990)}
\maketitle

\begin{abstract}
Quantum noise limits the sensitivity of interferometric measurements. It is
generally admitted that it leads to an ultimate sensitivity, the ``standard
quantum limit''. Using a semi-classical analysis of quantum noise, we show
that a judicious use of squeezed states allows one in principle to push the
sensitivity beyond this limit. This general method could be applied to large
scale interferometers designed for gravitational wave detection.

\end{abstract}

\begmc
Quantum noise ultimately limits the sensitivity in interferometric detection
of gravitational waves \cite{QLIM1,QLIM2,QLIM3}. A gravitational wave is
detected as a phase difference between the optical lengths of the two arms.
It seems accepted that there exists a ``standard quantum limit'' (SQL),
equivalent to an ultimate detectable length variation 
\begin{equation}
(\Delta z)_{SQL}=\sqrt{\frac{\hbar \tau }{M}}  \eqnum{1}
\end{equation}
where $M$ is the mass of the mirrors and $\tau $ the measurement time \cite
{QLIM4}. The SQL can be derived by considering that the positions $z(t)$ and 
$z(t+\tau )$, which are non commuting observables, are measured \cite{QLIM5}%
. This interpretation of SQL has given rise to a long controversy \cite
{QLIM6}.

Alternatively, the SQL can be understood by considering the quantum noise as
a sum of two contributions. Photon counting noise corresponds to
fluctuations of the number of photons detected in the two output ports while
radiation pressure noise stems from the random motion of the mirrors which
is sensitive to the fluctuations of the numbers of photons in each arm. The
sum of these two contributions leads to an optimal sensitivity given by the
expression (1). This limit is reached for very large laser power which is
not presently achievable.

Caves \cite{QLIM7} pointed out that these two contributions reflect the
fluctuations of two quadrature components of the vacuum field entering the
unused input port of the interferometer. He further suggested \cite{QLIM8} a
reduction in the photon counting noise by entering squeezed light \cite
{QLIM9} in this port. This possibility, which has been experimentally
demonstrated \cite{QLIM10}, allows one to attain the optimal sensitivity for
more reasonable laser powers. It does not overcome the SQL because the
reduction in photon noise is compensated by the increase in radiation
pressure fluctuations.

Unruh \cite{QLIM11} has shown that a judicious extension of Caves' proposal
leads to a sensitivity beyond the SQL. The key point is that photon counting
noise and radiation pressure noise are not independent sources of
fluctuations as implicitly assumed in the derivation of the SQL. Both
contributions are linearly superimposed in the fluctuations of the monitored
signal \cite{QLIM12}. It is therefore possible to reduce the total noise by
squeezing the appropriate quadrature component of the field entering the
unused input port.

We present in this letter a simple method for treating quantum noise in
interferometric measurements. It is based on a semi-classical linear input
output theory already used for computing the field fluctuations generated by
optical parametric oscillators \cite{QLIM13}. Using this general method, we
show that the sensitivity can be pushed beyond the SQL. The method can be
used for incorporating a detailed analysis of quantum noise in discussions
about large scale interferometers designed for gravitational wave detection.
As a first step in this direction, we discuss the quantum limits when taking
into account some constraints.

\section*{Measurement of the position of a mirror}

We first analyze the simple case where the position of a single mirror is
measured. The mirror is illuminated by an incident laser beam. The phase of
the reflected field at some arbitrary point contains the information about
the mirror position.

The incident electric field ${\bf E}(t)$ will be written as the sum of a
prescribed monochromatic field (frequency $\omega _{0}$ and wavevector $%
k_{0}=\frac{\omega _{0}}{c}$) and the quantum fluctuations of a
monodimensional scalar field (propagation along the $z$ direction with one
polarization only, but all possible frequencies) 
\begin{eqnarray}
{\bf E}(t) &=&\sqrt{\frac{\hbar \omega _{0}}{2\varepsilon _{0}c}}\left( {\bf %
u}(t)e^{ik_{0}z-i\omega _{0}t}+{\bf u}^{\dagger }(t)e^{-ik_{0}z+i\omega
_{0}t}\right)   \nonumber \\
&&  \eqnum{2a} \\
{\bf u}(t) &=&\frac{1}{2}\left( <{\bf p}>+\delta {\bf p}(t)+i\delta {\bf q}%
(t)\right)   \eqnum{2b} \\
\delta {\bf x}(t) &=&{\bf x}-\left\langle {\bf x}\right\rangle   \eqnum{2c}
\end{eqnarray}
The operators ${\bf p}$ and ${\bf q}$ represent respectively the amplitude
and phase quadrature components of the field.

We will use a semi-classical description of quantum fluctuations \cite
{QLIM13}. The classical random variables $p$ and $q$ are defined so that
they fit the symmetrically ordered momenta of the quantum fluctuations ${\bf %
p}$ and ${\bf q}$ 
\begin{equation}
\left\langle x(t) y(t^{\prime })\right\rangle =\frac{1}{2}\left\langle {\bf x}%
(t){\bf y}(t^{\prime })+{\bf y}(t^{\prime }){\bf x}(t)\right\rangle  
\eqnum{3}
\end{equation}
They are characterized by the spectra of stationary random variables 
\begin{equation}
S_{xy}(\Omega )={\int_{-\infty }^{+\infty }}dt\left\langle \delta
x(t_{0}+t)\delta y(t_{0})\right\rangle e^{i\Omega t}  \eqnum{4}
\end{equation}
The spectra obey a generalized Heisenberg inequality \cite{QLIM14} which can
be written (in the frequency domain $\Omega \ll \omega _{0}$) as 
\begin{equation}
S_{pp}(\Omega )S_{qq}(\Omega )-S_{pq}(\Omega )^{2}\succeq 1  \eqnum{5}
\end{equation}

Using these notations, we can define a normalized intensity $I$, measured as
a number of photons per unit time, and a phase $\varphi $ 
\begin{eqnarray}
u(t)&=&\frac{1}{2}\left( \left\langle p\right\rangle +\delta p(t)
+i\delta q(t)\right) \nonumber \\ 
&=&\sqrt{I+\delta I(t)}\exp (i\delta \varphi (t))  \eqnum{6}
\end{eqnarray}
In a linear treatment of the field fluctuations, one obtains 
\begin{eqnarray}
I &=&\frac{1}{4}\left\langle p\right\rangle ^{2},\quad \delta I(t)=\frac{1}{2%
}\left\langle p\right\rangle \delta p(t)=\sqrt{I}\delta p(t)  \eqnum{7} \\
&&\delta \varphi (t)=\frac{\delta q(t)}{\left\langle p\right\rangle }=\frac{%
\delta q(t)}{2\sqrt{I}}  \eqnum{8}
\end{eqnarray}

We can now discuss the effect of quantum noise for this position
measurement. The phase of the reflected beam can be written 
\begin{equation}
\varphi =2k_{0}z+\delta \varphi (t)  \eqnum{9}
\end{equation}
where $\delta \varphi (t)$ represents the incident phase fluctuations (8)
and where the mirror position $z$ depends on the intensity fluctuations (7),
due to radiation pressure force $2\hbar k_{0}\left| u(t)\right| ^{2}$. In
order to evaluate this term, we have to describe the response of the mirror
to the force. In a linear analysis, this response is described in the
frequency domain by a susceptibility function $\chi $.

Thus, $\frac{\widetilde{\varphi }(\Omega )}{2k_{0}}$ provides an estimator $%
\widetilde{z}(\Omega )$ for each frequency component of the position 
\begin{equation}
\widetilde{z}(\Omega )=\overline{z}(\Omega )+\delta \overline{z}(\Omega ) 
\eqnum{10}
\end{equation}
where $\overline{z}(\Omega )$ and $\delta \overline{z}(\Omega )$ correspond
to the signal and noise. The noise is a sum of three error terms associated
respectively with the incident phase fluctuations $\delta z_{pc}$
superimposed to the signal, the mirror displacement $\delta z_{rp}$ due to
the radiation pressure and the mirror displacement $\delta z_{ef}$ due to
extra fluctuations $\delta \overline{f}$ 
\begin{eqnarray}
\delta \overline{z}(\Omega ) &=&\delta \overline{z}_{pc}(\Omega )+\delta 
\overline{z}_{rp}(\Omega )+\delta \overline{z}_{ef}(\Omega )  \eqnum{11a} \\
\delta \overline{z}_{pc}(\Omega ) &=&\frac{\delta \overline{q}(\Omega )}{%
4k_{0}\sqrt{I}}  \eqnum{11b} \\
\delta \overline{z}_{rp}(\Omega ) &=&\chi (\Omega )2\hbar k_{0}\sqrt{I}%
\delta \overline{p}(\Omega )  \eqnum{11c} \\
\delta \overline{z}_{ef}(\Omega ) &=&\chi (\Omega )\delta \overline{f}%
(\Omega )  \eqnum{11d}
\end{eqnarray}

\section*{Interferometric measurement}

Up to now, we have not discussed a practical realisation of the phase
measurement. In fact, this is the role of the interferometer to transform
the phase signal into an intensity signal.

The interferometer can be schematized as consisting of two input ports $A$
and $B$, two output ports $C$ and $D$ and two internal paths 1 and 2. We
will consider here a simple configuration. A mean field is entered only into
the port $A$. The beam splitters have equal transmission and reflection
probabilities and the difference $J$ between the two output intensities is
measured around a working point where it is zero. One obtains in this case 
\begin{eqnarray}
\delta I &=&\delta I_{1}-\delta I_{2}=\sqrt{I_{A}}\delta p_{B}(t) 
\eqnum{12a} \\
J &=&I_{C}-I_{D}=I_{A}2k_{0}z+\sqrt{I_{A}}\delta q_{B}(t)  \eqnum{12b}
\end{eqnarray}
where $z$ is the difference between the optical lengths of the two arms
(treated as a small quantity) and where $\delta p_{B}$ and $\delta q_{B}$
represent the amplitude and phase fluctuations of the field entering the
port $B$ ($I_{A}$ is the mean intensity entering the port $A$).

If the two mirrors have the same susceptibility $\chi $, the differential
displacement z can be written 
\begin{equation}
\overline{z}(\Omega )= \chi (\Omega )\left( -M\Omega ^{2}\overline{s}(\Omega
)+2\hbar k_{0}\delta \overline{I}(\Omega )+\delta \overline{f}(\Omega
)\right)  \eqnum{13}
\end{equation}
where the force is the sum of three terms associated respectively with the
gravitational wave, the radiation pressure and the extra fluctuations. The
gravitational signal is measured as the variation $s$ of a distance between
two free falling mirrors.

A signal estimator $\widetilde{s}(\Omega )$ can be defined at each frequency 
\begin{equation}
\widetilde{s}(\Omega )=\frac{\overline{J}(\Omega )}{2k_{0}I_{A}}=-M\Omega
^{2}\chi (\Omega )\overline{s}(\Omega )+\delta \overline{s}(\Omega ) 
\eqnum{14}
\end{equation}
where the error is, as before, the sum of three terms 
\begin{eqnarray}
\delta \overline{s}(\Omega ) &=&\delta \overline{s}_{pc}(\Omega )+\delta 
\overline{s}_{rp}(\Omega )+\delta \overline{s}_{ef}(\Omega )  \eqnum{15a} \\
\delta \overline{s}_{pc}(\Omega ) &=&\frac{\delta \overline{q}_{B}(\Omega )}{%
2k_{0}\sqrt{I_{A}}}  \eqnum{15b} \\
\delta \overline{s}_{rp}(\Omega ) &=&\chi (\Omega )2\hbar k_{0}\sqrt{I_{A}}%
\delta \overline{p}_{B}(\Omega )  \eqnum{15c} \\
\delta \overline{s}_{ef}(\Omega ) &=&\chi (\Omega )\delta \overline{f}%
(\Omega )  \eqnum{15d}
\end{eqnarray}

Quantum fluctuations and extra fluctuations are independent, but $p$ and $q$
fluctuations can be correlated. One obtains from (15) the noise spectrum 
\begin{eqnarray}
S_{ss}(\Omega ) &=&\frac{S_{qq}(\Omega )}{4k_{0}^{2}I_{A}}+2\hbar \chi
_{R}(\Omega )S_{pq}(\Omega )  \eqnum{16} \\
&+&\left( \chi _{R}^{2}(\Omega )+\chi _{I}^{2}(\Omega )\right) \left( 4\hbar
^{2}k_{0}^{2}I_{A}S_{pp}(\Omega )+S_{ff}(\Omega )\right)   \nonumber
\end{eqnarray}
where $\chi _{R}(\Omega )$ and $\chi _{I}(\Omega )$ are the real and
imaginary parts of $\chi (\Omega )$ (the spectra $S_{qq}$, $S_{pq}$ and $%
S_{pp}$ refer to the input port $B$).

We will consider that the signal is measured through a filter characterized
by a function $G(\Omega )$ (maximum value 1 at the signal frequency $\Omega
_{S}$). The filtered noise is given by the integral 
\begin{equation}
\Delta s^{2}=2B\overline{S_{ss}}  \eqnum{17}
\end{equation}
where $B$ is the detection bandwidth 
\begin{equation}
2B={\int_{-\infty }^{+\infty }}\frac{d\Omega }{2\pi }G(\Omega )  \eqnum{18}
\end{equation}
and where $\overline{F}$ is the mean value of a function $F$ over the
normalized frequency distribution $\frac{G}{4\pi B}$.

When the mirrors are held in their equilibrium positions by damped harmonic
systems, the susceptibility is 
\begin{equation}
\chi (\Omega )=\frac{1}{M\left( \Omega _{M}^{2}-\Omega ^{2}-i\Gamma \Omega
\right) }  \eqnum{19}
\end{equation}
and the signal is accurately reproduced when the eigenfrequency $\Omega _{M}$
and the damping constant $\Gamma $ are smaller than the signal frequency $%
\Omega _{S}$. More general expressions of $\chi (\Omega )$, which are
studied for the large scale interferometers, must also obey \cite{QLIM2} 
\begin{equation}
-M\Omega ^{2}\chi (\Omega )\symbol{126}1\quad {\rm for}\quad \Omega
_{S}-2B<\Omega <\Omega _{S}+2B  \eqnum{20}
\end{equation}

\section*{Quantum limits}

In the following, we disregard the extra fluctuations and consider only the
quantum noise 
\begin{eqnarray}
\Delta s^{2} &=&\frac{2B}{4k_{0}^{2}I_{A}}\overline{S_{qq}}+4B\hbar 
\overline{\chi _{R}S_{pq}}  \nonumber \\
&&+8B\hbar ^{2}k_{0}^{2}I_{A}\overline{\left( \chi _{R}^{2}+\chi
_{I}^{2}\right) S_{pp}}  \eqnum{21}
\end{eqnarray}
We first consider the simplest case where the fluctuations entering the port 
$B$ are vacuum fluctuations ($S_{pp}(\Omega )=S_{qq}(\Omega )=1$; $%
S_{pq}(\Omega )=0$) 
\begin{equation}
\Delta s^{2}=\frac{2B}{4k_{0}^{2}I_{A}}+8B\hbar ^{2}k_{0}^{2}I_{A}
\overline{ \chi _{R}^{2}+\chi _{I}^{2} }  \eqnum{22}
\end{equation}
In this case, the quantum noise is effectively the sum of two independent
contributions associated with phase and amplitude fluctuations. Its minimum
when the laser intensity $I_{A}$ is varied is the standard quantum limit 
\begin{equation}
\Delta s_{SQL}^{2}=4B\hbar \sqrt{\overline{ \chi _{R}^{2}+\chi
_{I}^{2} }}  \eqnum{23}
\end{equation}
Using condition (20), we obtain the usual expression (1) with a time
parameter $\tau \approx \frac{4B}{\Omega _{S}^{2}}$.

Caves' proposal \cite{QLIM8} corresponds to squeezed phase fluctuations ($%
S_{pp}(\Omega )=K$; $S_{qq}(\Omega )=\frac{1}{K}$; $S_{pq}(\Omega )=0$). It
also leads to the standard quantum limit (23), but for a smaller laser
intensity.

We finally consider Unruh's proposal \cite{QLIM11} where correlated squeezed
fluctuations enter the input port $B$. The noise $\Delta s^{2}$ can be
decreased below the SQL by varying $S_{qq}(\Omega )$, $S_{pq}(\Omega )$ and $%
S_{pp}(\Omega )$ and respecting the Heisenberg inequality (5). A lower bound
for the sensitivity is obtained by assuming that the squeezing can be
optimised at each frequency. One finds in this way 
\begin{equation}
\Delta s^{2}=4B\hbar \overline{\left| \chi _{I}\right| }  \eqnum{24}
\end{equation}
This lower bound is far below the standard quantum limit (23) since the
reactive part $\chi _{R}$ of the susceptibility is much larger than the
dissipative part $\chi _{I}$ when the condition (20) is satisfied. For
example, the damped harmonic system leads to the expression (1) with a time
parameter $\tau \approx \frac{4B\Gamma }{\Omega _{S}^{3}}$. It has to be
noted that the recoil effect associated with the reflection of photons \cite
{QLIM11} gives rise to a damping $\Gamma _{min}=\frac{\hbar k_{0}I_{A}}{M}$.
In practice, the damping constant is larger than this minimum value but the
lower bound (24) is still below the SQL (23) as long as $\Gamma <\Omega _{S}$%
.

If broadband correlated squeezing is used, one has another minimum noise 
\begin{equation}
\Delta s^{2}=4B\hbar \sqrt{\overline{\chi _{R}^{2}+\chi _{I}^{2}}-\overline{%
\chi _{R}}^{2}}  \eqnum{25}
\end{equation}
This noise is intermediate between the SQL (23) and the lower bound (24). It
can be shown that it is close to the SQL (23) when condition (20) is
satisfied. This shows that it is important to control the squeezing
parameters at each frequency in order to approach the lower bound (24).

We are grateful to A.Heidmann, E.Giacobino and C.Fabre for stimulating
discussions.

\endmc


\begin{references}
\bibitem{QLIM1}  Meystre P. and Scully M.O. (eds) {\it Quantum Optics,
Experimental Gravitation and Measurement Theory} (Plenum, 1983)

\bibitem{QLIM2}  Brillet A., Damour T. and Tourrenc Ph., {\it Ann. Physique} 
{\bf 10} 201 (1985); Brillet A., {\it Ann. Physique} {\bf 10} 219 (1985)

\bibitem{QLIM3}  Gea-Banacloche J. and Leuchs G., {\it J. Opt. Soc. Am.} 
{\bf B4} 1667 (1987); {\it J. Mod. Opt.} {\bf 34} 793 (1987)

\bibitem{QLIM4}  Braginski V.B. and Vorontsov Yu.I., {\it Usp. Fiz. Nauk} 
{\bf 114} 41 (1974), {\it Sov. Phys. Usp.} {\bf 17} 644 (1975)

\bibitem{QLIM5}  Caves C.M., Thorne K.S., Drever R.W.P., Sandberg V.D. and
Zimmermann M, {\it Rev. Mod. Phys.} {\bf 52} 341 (1980); Caves C.M. in ref.
[1] p.567; Caves C.M. {\it Phys. Rev. Lett.} {\bf 54} 2465 (1985)

\bibitem{QLIM6}  Yuen H.P. {\it Phys. Rev. Lett.} {\bf 51} 719 (1983); Ozawa
M. {\it Phys. Rev. Lett.} {\bf 60} 385 (1988); {\it Phys. Rev.} {\bf A41}
1735 (1990)

\bibitem{QLIM7}  Caves C.M., {\it Phys. Rev. Lett.} {\bf 45} 75 (1980)

\bibitem{QLIM8}  Caves C.M., {\it Phys. Rev.} {\bf D23} 1693 (1981)

\bibitem{QLIM9}  General references on squeezing can be found in Loudon R.
and Knight P. (Eds) {\it Squeezed Light, J. Mod. Opt.} {\bf 34} 709-1020
(1987); Kimble H.J. and Walls D.F. (Eds) {\it Squeezed States of the
Electromagnetic Field, J. Opt. Soc. Am.} {\bf B4} 1449-1741 (1987); Tombesi
P. and Pike E.R. (Eds) {\it Squeezed and non classical light} (Plenum, 1989)

\bibitem{QLIM10}  Grangier Ph., Slusher R.E., Yurke B. and LaPorta A., {\it %
Phys. Rev. Lett.} {\bf 59} 2566 (1987); Min Xiao, Wu L.A. and Kimble H.J., 
{\it Phys. Rev. Lett.} {\bf 59} 2781 (1987)

\bibitem{QLIM11}  Unruh W.G. in ref. [1] p.647

\bibitem{QLIM12}  Loudon R., {\it Phys. Rev. Lett.} {\bf 47} 815 (1981);
Bondurant R.S., {\it Phys. Rev.} {\bf A34} 3927 (1986)

\bibitem{QLIM13}  Reynaud S., Fabre C. and Giacobino E., {\it J. Opt. Soc.
Am.} {\bf B4} 1520 (1987); Reynaud S. and Heidmann A., {\it Opt. Commun.} 
{\bf 71} 209 (1989); Reynaud S., {\it Ann. Physique} {\bf 15} 63 (1990)

\bibitem{QLIM14}  L\'{e}vy-Leblond J.M., {\it Am. J. Phys.} {\bf 54} 135
(1986); Schumaker B.L., {\it Phys. Rep.} {\bf 135} 317 (1986); Luks A.,
Perinova V. and Perina J., {\it Opt. Commun.} {\bf 67} 149 (1988)
\end{references}
\end{document}